\documentstyle[amssymb,secnumtab,epsf]{fbssuppl}

\title{QCD Sum Rules and Soft-Hard Interplay for Hadronic Form Factors}
\author{A. V. Radyushkin\thanks{{\it Also at:} Laboratory of 
Theoretical Physics, JINR, Dubna, Russia }}
\institute{Physics Department, Old Dominion University,
Norfolk, VA 23529, USA and Jefferson Lab, Newport News,VA 23606, USA}

\begin{document}

\maketitle
\begin{abstract}
We discuss  two  types of contributions
to hadronic form factors in QCD:  hard 
gluon exchange and soft wave  function overlap.
Within the QCD sum rule approach,
the hard contribution has strong 
numeric suppression by factor 
 $(\alpha_s/\pi)\sim 0.1$ 
for each exchange.
For this reason, the soft contribution 
dominates at accessible momentum transfers.
The  ``humpy'' distribution amplitudes 
used to enhance hard terms cannot be derived
from QCD sum rules in a self-consistent way. 
The estimates of soft terms obtained within
the local quark-hadronic duality approach 
in all cases are close to existing data,
providing an experimental evidence
that hard terms are small.
\end{abstract}

\section{Soft {\it vs.} Hard}

It is still a matter of controversy
whether  hard scattering \cite{bf} or 
 the soft wave function overlap mechanism 
 \cite{feynman}) 
is responsible  for the 
experimentally observed power-law behaviour
of elastic hadronic form factors. 
At sufficiently  large momentum transfer, 
the soft mechanism  is dominated  
by  configurations in which  one of 
the quarks carries  almost
all the momentum of the hadron.
On the other hand, the hard scattering term is 
generated by  the   
valence configurations with  small
transverse sizes and  finite light-cone fractions
of the  total hadron momentum
carried by each valence  quark. 
For large $Q^2$ in QCD,  this difference 
results in  an extra  $1/Q^2$-suppression 
of the soft term compared to the hard scattering one.

The hard term 
 can be written in a  factorized form \cite{cz},\cite{er},\cite{lb}
 as a product of a 
perturbatively calculable hard scattering amplitude and
two distribution amplitudes (DAs) 
describing how the large longitudinal
momentum of the  initial and final hadrons
is shared by their constituents.
This mechanism involves exchange 
of virtual gluons, each exchange 
 bringing in a  noticeable  suppression 
factor $(\alpha_s/{\pi}) \sim  0.1$.
As a  result,  to describe  existing data 
by the hard contribution alone,
one should   increase somehow 
the magnitude of the hard scattering term.

This is usually achieved  by using 
the DAs
with  a ``humped''  shape \cite{cz2}.
However, the   passive quarks in this situation 
carry  a rather small fraction of the hadron momentum and,
 as   pointed out in ref.\cite{isgur},  the ``hard'' scattering
subprocess,   even at  rather  large momentum transfers 
$Q^2\sim 10\, GeV^2$, is dominated  by  rather  small  gluon
 virtualities. 
This means that the hard scattering scenario heavily relies 
on the assumption that the asymptotic pQCD  
approximations ($e.g.,$ the $1/k^2$-behaviour 
of the gluon propagator $D^c(k)$)   are accurate even for 
momenta $k$ smaller than  $300 \, MeV$,
$i.e.,$ in the region strongly affected by 
finite-size effects, nonperturbative QCD vacuum fluctuations,
$etc.$   Including  these effects 
decreases the magnitude of the gluon 
propagator $D^c(k)$ at small spacelike $k$  converting 
$D^c(k)$ into something like $1/(k^2 - \Lambda^2)$ and 
shifts   the hard contributions significantly  
below the data level even if one uses 
the  humpy DAs and other modifications 
increasing  the hard term (see, $e.g.,$ \cite{kroll}).

 An instructive illustration of possible modifications
 due to finite size or transverse momentum effects 
 is given by the light-cone calculation 
 of the $\gamma^* \gamma \pi^0$ 
 amplitude \cite{bhl,murad} in which hard  
 propagator of a {\it massless}  
 quark is convoluted with the two-body wave function
 $\Psi (x,k_{\perp})$.
 Assuming the   Gaussian dependence \mbox{$\Psi (x,k_{\perp}) \sim
 \exp[ - k_{\perp}^2/2 x \bar x \sigma]$} on  transverse momentum,
 one can easily calculate the $k_{\perp}$ integral to see
   that the pQCD propagator factor $1/xQ^2$ is substituted
 by the combination $(1- \exp[ - xQ^2/2  \bar x \sigma])/xQ^2$
 which  monotonically 
 tends to a finite limit $1/2 \sigma$ as $x \to 0$.
 Hence, the effective virtuality is always larger than $2 \sigma$. 
 The suppression
 of low virtualities has a simple 
 explanation: propagation of quarks and gluons in the 
 transverse direction 
 is restricted by the finite size of the hadron. 
Numerically, 
$2 \sigma \approx 1.35 \, {\rm GeV}^2$ in that case.   
 However, even a milder  modification of  the ``hard''  
 propagators by 
  effective quark and gluon 
 masses $1/k^2 \to 1/(k^2 - M^2)$ with 
 $M^2 \sim 0.1 \, {\rm GeV}^2$ or model 
  inclusion of transverse momentum effects
 strongly reduces the magnitude of  hard contributions
 \cite{kroll}, especially
 when the humpy DAs are used.
   For these reasons,  a  scenario 
 with humpy DAs and bare $\sim 1/x_iy_jt$ propagators  
 (which amounts to ignoring finite-size effects)   
considerably overestimates   the size  of hard contributions. 
   
\section{Lessons from Pion Studies} 

   The  relative smallness of  hard contributions
   can be easily understood within the QCD sum rule 
   context. The soft contribution  is dual to the
   lowest-order diagram while the gluon exchange 
   terms appear in diagrams having a higher order 
   in $\alpha_s$ which results in the usual 
   $\alpha_s/\pi \sim 1/10$  suppression factor 
   per each extra loop. 
   In particular, the $\alpha_s/\pi $ suppression factor is 
   clearly visible in the expression for the hard contribution
   to the pion form factor \cite{czs,farjack,er,bl} 
\begin{equation}    
   F_{\pi}^{\rm hard}(Q^2) |_{\varphi_{\pi} = \varphi_{\pi}^{as}}
   = \frac{8 \pi \alpha_s f_{\pi}^2}{Q^2} = 2 
   \left ( \frac{\alpha_s}{ \pi} \right  )  \frac{s_0}{Q^2} \, .
  \end{equation} 
Here, the combination $s_0 = 4 \pi^2 f_{\pi}^2 
\approx 0.67$ GeV$^2 \sim m_{\rho}^2 $ 
is what is usually called the  ``typical  
hadronic  scale'' in the case of the pion.   
   At asymptotically high
   $Q^2$, the  $O(\alpha_s/\pi)$
    suppression of the hard terms is more than compensated
   by their slower decrease with $Q^2$. 
   However, such a compensation does not occur
   in the subasymptotic region where  
   the soft contributions, as we have seen, may 
    have the same effective  power 
   behavior as that  predicted by the asymptotic
   quark counting rules for the hard contributions.
    In  ref. \cite{sjir},  both  the soft contribution
   and the $O(\alpha_s)$ corrections for the pion form factor 
   were calculated together within a QCD sum rule
   inspired  approach.
   The ratio of the $O(\alpha_s)$ terms to the 
   soft contribution  was shown to be in full agreement 
   with the expectation based on the 
    $\alpha_s/\pi $ per loop suppression.

The use of the  humpy DAs is usually motivated by
the  QCD sum rule analysis for the $\langle x^N \rangle $ moments 
of DAs \cite{cz2}.
However,  applications of the QCD sum rules to
 DAs $\varphi(x)$ and form factors $F(Q^2)$
require a more detailed information 
about the nonperturbative QCD vacuum  than 
those for the simpler classic cases \cite{SVZ}  of  hadronic masses 
and decay widths. 
The main problem 
is that the coefficients 
of the operator product expansion 
(OPE) for the relevant correlators
now  depends  on  an  extra parameter, $e.g.,$ on the 
order  of the moment $N$ for
$\langle x^N \rangle$ or momentum transfer  $Q^2$
for form factors. 
In particular, the higher  condensates   
$\langle \bar q (D^2)^n q \rangle$ 
are accompanied  in the $\langle x^N \rangle $ sum rule 
by large $N^n$-factors.   
Since for any reasonable shape of 
$\varphi(x)$  the moments  $\langle x^N \rangle $ 
should decrease with growing $N$,
the appearance of  $N^n$-dependence is an artifact 
of the expansion procedure.  
Calculationally, the $N^n$-factors  appear from the 
Taylor expansion of the nonlocal  condensate
 $\langle \bar q(0) q(z) \rangle$. 
In this situation, one is forced to make, explicitly or implicitly
an assumption about the structure  of the OPE in higher terms.
In  the  approach of ref.\cite{cz2}  only 
 the  lowest condensates were taken into account.
A simple   alternative  is to  model
$\langle \bar q(0) q(z) \rangle$ 
by a smooth function with the width suggested by 
existing estimates $\langle \bar q D^2 q \rangle / 
\langle \bar q  q \rangle \approx 0.4 \, GeV^2$.
This model  gives  a  QCD sum rule 
in which all terms decrease  for large $N$.
It produces the pion    DA
close to a smooth ``asymptotic'' form (see \cite{mr}). 

It was also   observed   that  the sum rules with nonlocal 
condensates have the property that the 
humps  in the relevant  correlator 
(corresponding to a sum over all possible states) 
get more pronounced  when 
 the relative pion contribution $decreases$ (see ref.\cite{minn}). 
This means  that the humps of the correlator
are generated by  oscillations in the DAs of the 
 higher states rather than by the humps in the pion DA. 
The oscillatory behaviour of DAs  of the radial 
excitations found in  ref.\cite{minn} (see also \cite{mikhbak}) 
is  supported by the studies in 
two-dimensional QCD \cite{qcd2,ericqcd2}.

An independent  evidence in favour 
of the narrow form of the
pion distribution amplitude 
 $\varphi_{\pi}(x)$ is provided by the result
of ref.\cite{braunfil}, where it was found
that  $\varphi_{\pi}(1/2) \approx 1.2 f_{\pi}$, to be compared
with $\varphi_{\pi}^{as}(1/2) = 1.5f_{\pi}$ for 
the asymptotic distribution amplitude \cite{er},\cite{lb}
and $\varphi_{\pi}^{CZ}(1/2) =0$
for the  CZ form \cite{cz}. 
Furthermore, the lattice calculation 
of ref.\cite{gupta} gives a rather  small value  
$\langle \xi^2 \rangle \approx 0.11$
for the second moment of the pion DA incompatible
with the humpy form (compare
with $\langle \xi^2 \rangle^{CZ} =0.43$
and $\langle \xi^2 \rangle^{as} =0.2$).
The statement that the pion DA is close 
to its asymptotic form even at a low normalization point
is also supported     
by calculation of the
pion DA in the chiral soliton model \cite{petpob} 
and by a direct QCD sum rule calculation of the 
large-$Q^2$ behavior of the 
$\gamma^* \gamma \pi^0$ form factor \cite{rrnp}.
Within the  light-cone QCD sum rule approach 
one can relate the pion DA to the pion parton densities \cite{beljo}
known experimentally.
According to the analysis performed in \cite{beljo2},
existing data  favor the  asymptotic shape. 
Finally, 
the humpy  pion DA  advocated  in 
\cite{czpi,cz}  
is now ruled out by recent experimental data \cite{cleo} 
on the $\gamma^* \gamma \pi^0$ form factor. 
The data  are fully consistent with the next-to-leading 
pQCD prediction  calculated using  the asymptotic DA
\cite{braaten,murad,brojipar}.

If the pion   DA  is narrow,
the hard contribution to  the pion form factor  is small.   
On the other hand, in many models, the soft term  calculated as 
 an overlap of  model  wave functions 
$\Psi(x, k_{\perp})$ is  comparable in size with
the data \cite{isgur,kissl,cotanch,ericff}.  It should be noted 
that the relevant distribution amplitudes 
(obtained from $\Psi(x, k_{\perp})$  by integration over
$k_{\perp}$),   are narrow and the 
hard term is small. 
Moreover,  if one intends to increase  the hard term by 
using  wave functions enhanced in the end-point regions,
one also increases  the soft term (see, $e.g.$, \cite{isgur,ericff}),
since the latter is dominated by the 
regions where one of the quarks has small momentum.

The pion form factor was also studied within the 
QCD sum rule approach, which 
 is applicable in that case both in the region of moderately large 
\cite{i1},\cite{nr82} and  small momentum transfers  \cite{nr84}.
In the whole region 
$0  \leq Q^2  \raisebox{-.2ex}{$\stackrel{\textstyle<}
{\raisebox{-.6ex}[0ex][0ex]{$\sim$}}$} \, 3 \, GeV^2$,
the  QCD sum rule result for the contribution due to the 
Feynman mechanism is sufficiently large  to explain 
the magnitude of  existing data.
 In the region $Q^2 \raisebox{-.2ex}{$\stackrel{\textstyle>}
{\raisebox{-.6ex}[0ex][0ex]{$\sim$}}$} \, 4 \, GeV^2$,  the OPE  is ruined 
by  $O(Q^2/M^2)^n$  enhancement of condensate contributions.
This phenomenon  has exactly the same nature as the 
 $O(N)$  enhancement of the lowest condensate contributions
  in the CZ  sum rule  for $\langle x^N \rangle$ \cite{bar92}.
As a result,  if we  would assume  that
the higher condensate corrections can be neglected, 
we  would get a very large 
soft contribution, marginally exceeding the data.  
Alternatively, using the nonlocal condensates \cite{mr} 
 we would get  the soft term  comparable in size 
with the data \cite{br}. 
As mentioned above, the same model produces  a 
rather narrow pion DA \cite{mr}  generating 
a   small hard contribution which is 
subdominant    up to  $Q^2  \sim 10 \, GeV^2$. 
Since the  QCD sum rules for  $\langle x^N \rangle $ 
and $F^{soft}(Q^2)$ have similar structure, 
if one uses the same model for the
 condensates in both sum rules, 
the results for the hard term $F^{hard}(Q^2)$ 
(whose magnitude is determined 
by the shape of $\varphi(x)$ )
and soft term $F^{soft}(Q^2)$ 
are strongly correlated. 
Just like in quark model calculations, 
it is impossible to get a large hard term  
without getting a huge soft term. 
 The existence of such a  correlation is also supported  by 
the  light-cone QCD sum rules \cite{braunhal}.
Just like in quark model calculations, 
it is impossible to get a large hard term  
without getting a huge soft term.

 \section{Nucleon Case} 

Since  the structure of OPE in the pion and 
nucleon cases is  very similar,  
there is no reason   
to expect  a significant  deviation of the nucleon DA  
from its asymptotic form. 
In particular, an evidence against humpy nucleon 
DAs is provided by a lattice calculation \cite{marsac} 
which does not indicate any significant asymmetry. 
One may argue 
that the proton DA  must be asymmetric to  
reflect  the  fact that
the $u$-quarks carry on average a larger  fraction 
of the proton momentum than the $d$-quarks. 
As shown in ref. \cite{bokroll},   to 
accomodate this observation one needs only 
a moderate shift of the DA maximum from the center point
$x_1=x_2=x_3=1/3$.
 Such a shift does not produce a  drastic enhancement
 of the hard contribution provided by the humpy  DAs.
 However, with the asymptotic DA,
the leading twist hard contribution  completely 
fails to describe the data: 
 it gives   zero for the proton magnetic form factor
and a wrong-sign (positive)  contribution for the neutron
magnetic form factor,
with the  absolute magnitude of the latter being two orders 
of magnitude below the data \cite{belioffe}.

In the case of the baryon form factors, 
the standard SVZ-Borel version of the QCD 
sum rule approach works only in the region 
of small momentum transfers $Q^2 \raisebox{-.2ex}
{$\stackrel{\textstyle<}{\raisebox{-.6ex}[0ex][0ex]{$\sim$}}$} 1 \,  GeV^2$ 
\cite{belkogan}.
Beyond this region,  the OPE  explodes because of 
 $O(Q^2/M^2)$-enhancements in  
condensate  contributions, and a regular QCD sum rule analysis
is impossible. 
In ref.\cite{nr83}, it was proposed  to  
estimate the soft contributions  
by using  the local quark-hadron duality prescription.  
It amounts to  calculating 
  the amplitude for   transitions between the 
free-quark states produced  
(or annihilated) by a local current 
having the  hadron's quantum numbers, 
and then  averaging  
the invariant mass of the quark states 
over the appropriate duality interval $s_0$.
The latter  has the meaning of the effective threshold 
for the higher hadronic states in the relevant channel and  
has a specific value for each 
hadron, $e.g.,$ $s_0^{\pi} \approx 0.7 \, GeV^2$ 
for the pion in the axial current channel.

For the pion form factor, the  local quark-hadron duality
is supported by the QCD sum rule analysis \cite{i1,nr82}
and  agrees well with experimental data.
Furthermore, as argued in ref. \cite{nr83,apa} 
(see also \cite{qhd95})
the  quark-hadron duality  prescription 
has an intuitively appealing 
interpretation in terms of the light-cone 
wave functions: it can be treated as a cut-off model 
$\Psi( \{x_i\}, \{k_{\perp_i}\}) \sim 
\theta( \sum_i (k_{\perp_i}^2 /x_i) \leq s_0)$ 
 for the  soft wave function.  
The sharp  cut-off   suggested by the local duality 
looks like a rough approximation for more 
smooth wave functions  usually adopted in phenomenological 
quark models. However,  the difference is
that in  the local duality 
model the width of the  $k_{\perp}$-distribution
is directly  related to a parameter $s_0$ 
characterizing the  hadronic spectrum.
This parameter  $s_0$ is 
calculated  from the reliable two-point
function QCD sum  rule and considered as given
in the form factor calculations.

The  local duality  estimate \cite{nr83} of the 
soft term for the proton magnetic form factor,   
based on the standard value  
$s_0^{N} \approx 2.3 \, GeV^2$ \cite{belioffe} 
of  the nucleon duality interval 
is very close to  available data \cite{slac36},
\cite{slac883} over a wide region
$3 \, GeV^2 \raisebox{-.2ex}{$\stackrel{\textstyle<}
{\raisebox{-.6ex}[0ex][0ex]{$\sim$}}$} Q^2 \, 
\raisebox{-.2ex}{$\stackrel{\textstyle<}
{\raisebox{-.6ex}[0ex][0ex]{$\sim$}}$} \,  20\, GeV^2$.
The same calculation  \cite{nr83} also 
 correctly reproduces the observed magnitude 
 of the helicity-nonconservation  effects 
for the proton form factors: 
$ F_2^p(Q^2)/F_1^p(Q^2) \sim \mu^2/Q^2 $ 
with $\mu^2 \sim 1 \, GeV^2 $ \cite{slac883}.
 Within the scenario based on  hard scattering dominance,
it is  rather difficult to understand 
the origin of such a large scale,
since possible sources of  helicity nonconservation 
 in pQCD include  
only small scales like quark masses,
intrinsic transverse momenta, $etc.$, and 
one would rather expect that  
$ \mu^2 \sim 0.1 \, GeV^2$.

\section{Proton to Delta Transition}

Even more drastic difference between 
predictions of hard and soft scenarios 
is expected (see, $e.g.,$ \cite{burkert95})  
in the studies of  spin effects in the 
$\gamma^* p \to \Delta^+$ transition.
A renewed  attention to this process was 
raised by the results \cite{stoler1} of the analysis of inclusive 
SLAC  data which indicated that the effective transition 
form factor drops faster than one would expect from 
quark counting rules \cite{bf,mmt,bldelta}.
Within the  hard scattering scenario, the DA-sensitivity  of 
this process  was originally analyzed in ref.\cite{carlson}.
It was observed there that
the hard scattering amplitude in this case
has an extra suppression due to cancellation 
between  symmetric and antisymmetric parts 
of the nucleon distribution amplitude.
Hence,  from the hard scenario point of view,
the faster fall-off
found in \cite{stoler1} signalizes   the dominance 
of a non-asymptotic  contribution.

In ref.\cite{delta}, the soft contribution
to the $\gamma^* p \to \Delta^+$ transition
form factors was estimated within the local quark-hadron duality 
approach.  The duality interval
for the $\Delta$-resonance  taken there 
is $s_0^{\Delta} = 3.5$ GeV$^2$,
which agrees with the results of the
two-point function analysis \cite{belioffe}.
The results  for the  effective form factor
$G_T(Q^2)$ are  close to
those obtained from the 
analysis of inclusive  data \cite{slac}.
This means that the data can be  described without 
a sizable contribution from the  
hard-scattering mechanism. 
Furthermore, the $\gamma^* p \to \Delta^+$ transition
is described by three independent form factors,
and  a correct  model should not only be able to 
adjust the absolute magnitude of one of them:
it   should also be  able to explain  the relations 
between different form factors.
In particular,  the pQCD calculation  \cite{carlson}
predicts  that the lowest-twist hard
 contribution always has the property
$G_E^{* \, hard}(Q^2) \approx - G_M^{* \, hard} (Q^2)$.
This prediction is a specific example 
of the helicity selection rules \cite{lb}
inherent in  the hard scattering mechanism.
Experimentally, the ratio $G_E^*(Q^2)/G_M^*(Q^2)$
is very small \cite{burkert95,new}, which indicates that 
the leading-twist pQCD term is irrelevant 
in the region $Q^2 \raisebox{-.2ex}{$\stackrel{\textstyle<}
{\raisebox{-.6ex}[0ex][0ex]{$\sim$}}$} \, 4 \, GeV^2$.
Small value for $G_E^*(Q^2)/G_M^*(Q^2)$ is 
also predicted in constituent quark model approaches
\cite{cqmisgur,cqmclose,cqmcarl,cqmroll,cqminna}.
However, these approaches usually 
do not claim applicability 
in  the $Q^2 \gtrsim \, 2 \, $ GeV$^2$ region of momentum transfers. 
 The local duality estimates were performed
 in ref. \cite{delta}
for  several  Lorentz structures  which 
appear in the decomposition  of the basic
$\gamma$-odd three-point amplitude. 
The results
obtained from different invariant amplitudes are in satisfactory agreement
with each other.
All  estimates  indicate 
that the transition is dominated by the magnetic form factor
$G_M^*(Q^2)$, with electric $G_E^*(Q^2)$
and Coulomb  $G_C^*(Q^2)$ form factors being 
small compared to $G_M^*(Q^2)$ for all experimentally accessible
momentum transfers (see Fig. 1).

To summarize,  QCD sum rule based results
for   soft contributions to hadronic form factors 
are in good quantitative agreement with existing data
providing a clear experimental evidence that 
at available $Q^2$ hard terms are relatively small.

\begin{figure}[htb]
\mbox{\hspace{-1cm} \epsfxsize=7cm
  \epsfysize=5cm \epsffile{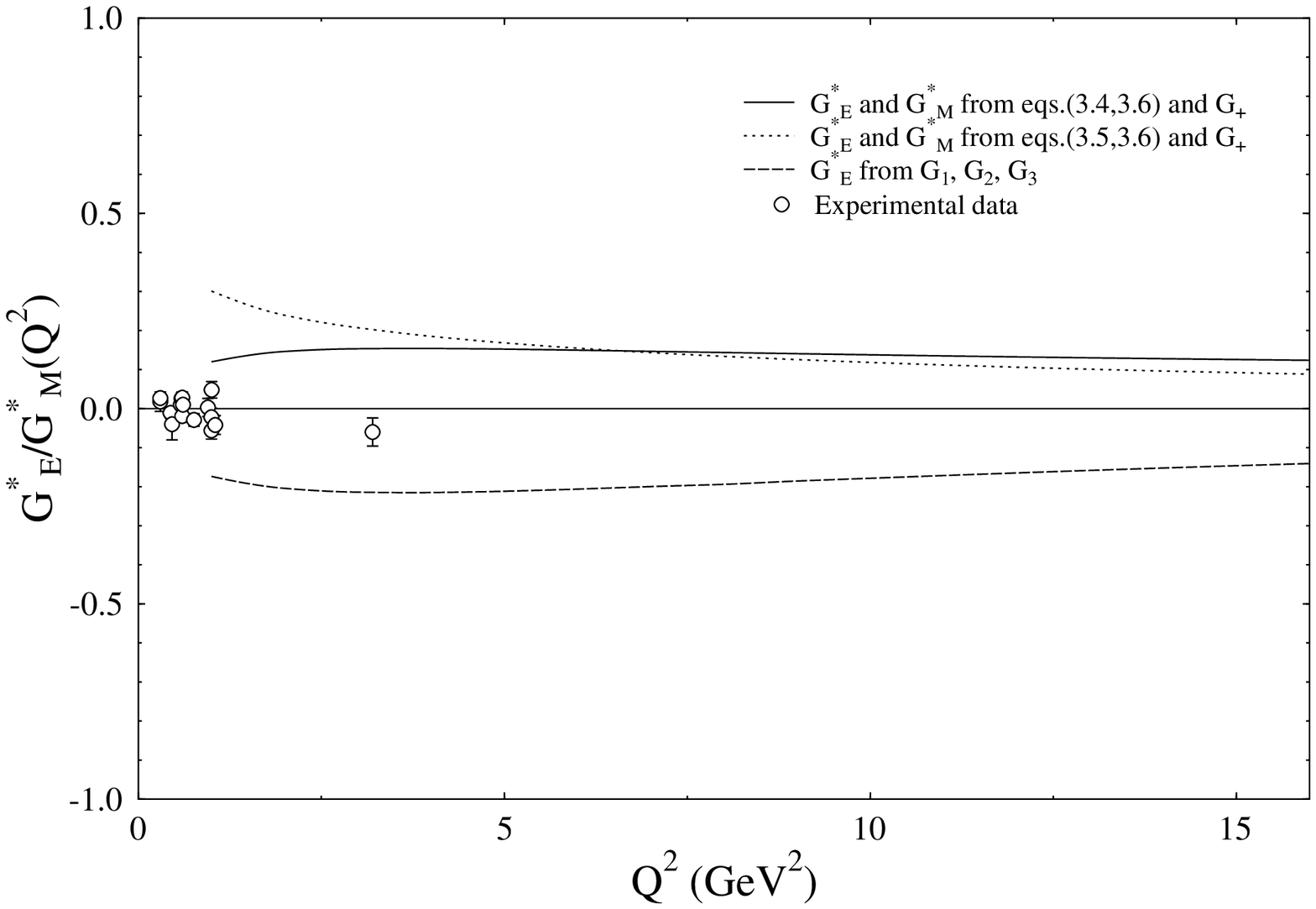}  \epsfxsize=7cm
  \epsfysize=5cm \epsffile{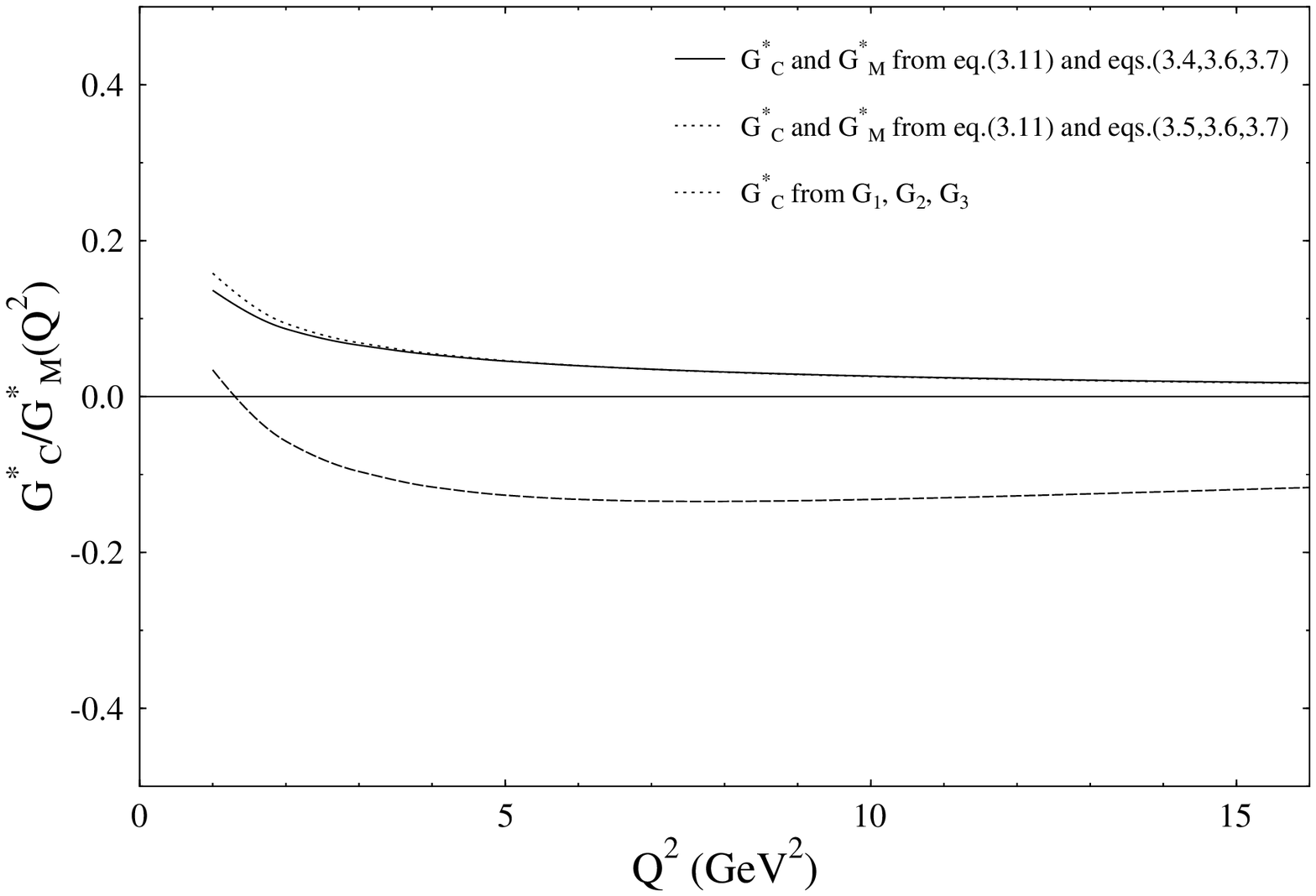}}
 \vspace{-1cm}
{\caption{\label{fig:4}   Local duality 
estimates for  the ratio of form factors {\it a)}
$G_E^*(Q^2)$ and $G_M^*(Q^2)$ and {\it b)} $G_C^*(Q^2)/G_M^*(Q^2)$.
   }}
\end{figure}

{\it Acknowledgement.} I am   grateful 
to I.G. Aznauryan, V.M. Belyaev, V.D. Burkert, C.E. Carlson, 
V.V. Frolov, N. Isgur, C. Keppel,
 N. Mukhopadhyay 
and P. Stoler for useful 
discussions.
This work was supported  by the US Department of Energy under 
contract DE-AC05-84ER40150.


\begin{thebibliography}{99}

\bibitem{bf} S.J. Brodsky and G.R. Farrar: Phys. 
Rev. Lett. {\bf 31}, 1153 (1973)
\bibitem{feynman} R.P. Feynman: {\it Photon-Hadron Interaction}. 
Reading: W.A.Benjamin
1972
\bibitem{cz} V.L. Chernyak and A.R. Zhitnitsky: JETP Lett. {\bf 25},
 510 (1977); Yad. Fiz. {\bf 31},  1053 (1980)
\bibitem{er} A.V. Efremov and A.V. Radyushkin: Phys. Lett. {\bf B94}, 
 245 (1980); Theor. Mat. Fiz. {\bf 42}, 147 (1980) 
\bibitem{lb} G.P. Lepage and S.J. Brodsky: Phys. Lett. {\bf B87},
359 (1979); Phys. Rev. {\bf D22}, 2157 (1980)
\bibitem{cz2} V.L. Chernyak and A.R. Zhitnitsky: Phys. Rep. {\bf 112},
 173 (1984)
\bibitem{isgur} N. Isgur and C.H. Llewellyn-Smith: Nucl. Phys. {\bf B317},
  526 (1989)
\bibitem{kroll}  J. Bolz et al.:  Z.Phys. {\bf C66},  267 (1995)
\bibitem{bhl}  S.J. Brodsky, T.Huang  and G.P.Lepage: 
 In: {\it Particles and Fields 2}, Proceedings of the 
Banff Summer Institute, Banff, Alberta, 1981,
edited by A.Z. Capri and A.N. Kamal, p.143. New York: Plenum 1983. 
\bibitem{murad} I.V. Musatov and A.V. Radyushkin: 
 Phys. Rev. {\bf D56}, 2713  (1997)
  \bibitem{czs} V.L. Chernyak, A.R. Zhitnitsky and V.G. Serbo:
  JETP Lett. {\bf 26}, 594 (1977)
  \bibitem{farjack} G.R. Farrar and D.R. Jackson: 
  Phys. Rev.  Lett. {\bf 43 }, 246 (1979)
  \bibitem{bl}  G.P. Lepage and S.J. Brodsky: 
  Phys. Lett.  {\bf  B87}, 359 (1979) 
  \bibitem{sjir} A. Szczepaniak, C.-R. Ji and 
  A. Radyushkin: Phys.Rev.  {\bf D57}, 2813 (1998)
 \bibitem{SVZ}  M.A. Shifman, A.I. Vainshtein and V.I. Zakharov: Nucl. Phys.
{\bf B147}, 385, 448 (1979)
\bibitem{mr}  S.V. Mikhailov and A.V. Radyushkin: Phys. Rev. {\bf D45},
 1754 (1992)
\bibitem{minn} A.V. Radyushkin: In: ``Continuous advances in QCD'',
ed. by A.V.Smilga, p. 238. Singapore: World Scientific 1994;
hep-ph/9406237
\bibitem{mikhbak} A.P. Bakulev and S.V. Mikhailov: 
Z.Phys.  {\bf C65}, 451 (1995)
\bibitem{qcd2}    G.'t Hooft:  
Nucl. Phys. {\bf B75}, 461 (1974) 
\bibitem{ericqcd2}    B. Chibisov and A. R. Zhitnitsky:  
Phys.Lett. {\bf B362} 105 (1995) 
\bibitem{braunfil} V.M. Braun and I.Filyanov: 
Z. Phys. {\bf C44} 157 (1989)
\bibitem{gupta} D.Daniel, R.Gupta and D.G.Richards: Phys.  
Rev. {\bf D43} 3715  (1991) 
\bibitem{petpob} V.Yu. Petrov and  P.V. Pobylitsa:
  hep-ph/9712203
 \bibitem{rrnp} A.V. Radyushkin and R. Ruskov:   
Nucl. Phys.  {\bf B 481}, 625 (1996) 
 \bibitem{beljo} V.M. Belyaev and M. B. Johnson:  
   Phys.Rev.  {\bf D56}, 1481 (1997) 
   \bibitem{beljo2} V.M. Belyaev and M. B. Johnson: hep-ph/9703244
 \bibitem{czpi} V.L. Chernyak and A.R. Zhitnitsky:
Nucl. Phys.  {\bf  B201}, 492 (1984); Erratum:  {\bf  B 214}, 547  (1984)
 \bibitem{cleo} CLEO collaboration (J. Gronberg et al.): 
 {  Phys. Rev.  }  {\bf D57}, 33  (1998)
 \bibitem{braaten} E. Braaten,  Phys. Rev. {\bf D28}, 524  (1983)
 \bibitem{brojipar} S. J.  Brodsky,  C.-R. Ji, A. Pang and 
D. G. Robertson:  Phys.Rev.  {\bf D57}, 245 (1998) 
\bibitem{kissl}  O.C. Jacob and  L.S. Kisslinger: 
Phys. Lett. {\bf B243}, 323 (1990) 
\bibitem{cotanch} C.R. Ji, P.L. Chung and  S.R. Cotanch: 
Phys. Rev. {\bf D45}, 4214 (1992) 
\bibitem{ericff}   B. Chibisov and A. R. Zhitnitsky:  Phys.Rev.  
{\bf D52},  5273 (1995) 
\bibitem{i1} B.L. Ioffe and A.V. Smilga: Phys. Lett. {\bf B114},  353 (1982) 
\bibitem{nr82} V.A. Nesterenko and A.V. Radyushkin: Phys. Lett. {\bf B115},
410 (1982) 
\bibitem{nr84} V.A. Nesterenko and A.V. Radyushkin: JETP. Lett. {\bf 39},
 707 (1984) 
\bibitem{bar92} A.V. Radyushkin: In: ``Baryons `92'', ed. by M.Gai, p.366.
Singapore: World Scientific 1993
\bibitem{br} A.P. Bakulev and A.V. Radyushkin: Phys. Lett. {\bf B271}, 
223 (1991)
\bibitem{braunhal} V.M. Braun and I. Halperin:
 Phys. Lett. {\bf B328},  457 (1994) 
 \bibitem{marsac} G. Martinelli and C.T. Sachrajda:
Phys. Lett.  {\bf B217}, 319 (1989)
 \bibitem{bokroll} J. Bolz and  P. Kroll: 
 Z.Phys.  {\bf  A356}, 327 (1996)
 \bibitem{belioffe} V.M. Belyaev and B.L. Ioffe: Sov.Phys.JETP {\bf 56},
493 (1982).
 \bibitem{belkogan} V.M. Belyaev and I.I. Kogan: 
 Int.J.Mod.Phys. {\bf
 A8}, 153 (1993)
\bibitem{nr83} V.A. Nesterenko and A.V. Radyushkin: Phys. Lett. {\bf 128B}, 
 439 (1983); \\ Sov. J. Nucl. Phys.  {\bf  39}, 811 (1984)
\bibitem{apa} A.V. Radyushkin: Acta Phys. Pol. {\bf B15},   403 (1984) 
\bibitem{qhd95} A.V. Radyushkin:  Acta Phys. Pol.{\bf B26},  2067 (1995) 
\bibitem{slac36}  A.F. Sill et al.:  Phys. Rev.  {\bf D48}, 29  (1993) 
\bibitem{slac883} P. Bosted et al.: Phys. Rev. Lett. {\bf 68},
3841 (1992);
 L. Andivahis et al.: Phys.Rev. {\bf D50}, 5491 (1994) 
\bibitem{burkert95} V.D. Burkert and L. Elouadrhiri:
 Phys. Rev.  Lett. {\bf 75}, 3614 (1995)
\bibitem{stoler1} P. Stoler: Phys. Rev. Lett. {\bf 66},
 1003 (1991); Phys. Rev. {\bf D44},  73 (1991);
Phys. Reports {\bf 226},  103 (1993)
\bibitem{mmt} V.A.Matveev, R.M.Muradyan and A.N. Tavkhelidze:
Lett. Nuovo Cim. {\bf 7},  719 (1973)
\bibitem{bldelta} B.L.Ioffe:  Phys. Lett. {\bf 63B}, 
 425 (1976) 
\bibitem{carlson} C.E. Carlson: Phys.Rev. {\bf D34}, 2704 (1986) 
\bibitem{delta} V.M. Belyaev and A.V. Radyushkin,
Phys.Rev. {\bf D53}, 6509 (1996)
\bibitem{slac} L.M. Stuart et al.: Phys.Rev. {\bf D58}, 032003 (1998)  
\bibitem{new} V.V. Frolov et al.: hep-ex/9808024 
\bibitem{cqmisgur} R.Koniuk and N.Isgur:  
Phys. Rev.  Lett. {\bf 44},   485 (1980);
  Phys. Rev.  {\bf D21},   1868 (1980)
\bibitem{cqmclose}  Z.Li and F.E.Close:  Phys. Rev.  {\bf D42}, 2207 (1990)
\bibitem{cqmcarl} S. Capstick and G.Karl: Phys. Rev.  {\bf D41}, 2767 (1990);
S. Capstick, Phys. Rev.  {\bf D46},  2864 (1992) 
\bibitem{cqmroll}  M. Warns, H. Schroeder, W.Pfeil and H. Rollnik:
Z.Phys.  {\bf C45},  627 (1990)
\bibitem{cqminna}  I.G. Aznaurian:  Phys. Lett. {\bf B316}, 
 391 (1993); Z.Phys.  {\bf A346},  297 (1993) 






\end{thebibliography}
\end{document}